\documentclass[sigconf]{acmart}

\copyrightyear{2025}
\acmYear{2025}
\setcopyright{licensedusgovmixed}\acmConference[KDD'25]{Proceedings of the 31st ACM SIGKDD Conference on Knowledge Discovery and Data Mining V.2}{August 3--7, 2025}{Toronto, ON, Canada}
\acmBooktitle{Proceedings of the 31st ACM SIGKDD Conference on Knowledge Discovery and Data Mining V.2 (KDD'25). August 3--7, 2025, Toronto, ON, Canada}
\acmDOI{10.1145/3711896.3737431}
\acmISBN{979-8-4007-1454-2/2025/08}

\usepackage{times} 
\usepackage{helvet}
\usepackage{courier} 
\usepackage{graphicx}
\usepackage{natbib}
\usepackage{caption} 
\usepackage{todonotes}
\usepackage{booktabs}
\DeclareCaptionStyle{ruled}{labelfont=normalfont,labelsep=colon,strut=off}
\usepackage{algorithm}
\usepackage{algorithmic}
\usepackage{newfloat}
\usepackage{listings}
\usepackage{multicol}

\usepackage{bbding}
\usepackage{pifont}
\usepackage{wasysym}

\usepackage[frozencache,cachedir=.]{minted}

\usepackage{hyperref}
\hypersetup{
    colorlinks=true,
    linkcolor=blue,
    filecolor=magenta,      
    urlcolor=cyan,
    pdftitle={Overleaf Example},
    pdfpagemode=FullScreen,
    }
\floatstyle{ruled}
\newfloat{listing}{tb}{lst}{}
\floatname{listing}{Listing}

\definecolor{hyrumcolor}{RGB}{17, 227, 200}

\begin{document}

\title{EMBER2024 - A Benchmark Dataset for Holistic Evaluation of Malware Classifiers}

\author{Robert J. Joyce}
\orcid{0009-0003-7168-1237}
\affiliation{%
  \institution{Booz Allen Hamilton}
  \city{McLean}
  \state{VA}
  \country{USA} 
}
\email{Joyce_Robert2@bah.com}

\author{Gideon Miller}
\orcid{0009-0002-2561-3535}
\affiliation{%
  \institution{Laboratory for Physical Sciences}
  \city{College Park}
    \state{MD}
  \country{USA} 
}
\email{gmmiller@lps.umd.edu}

\author{Phil Roth}
\orcid{0009-0008-5404-4320}
\affiliation{%
  \institution{CrowdStrike}
  \city{Austin}
  \state{TX}
  \country{USA} 
}
\email{phil.roth@crowdstrike.com}

\author{Richard Zak}
\orcid{0000-0003-4272-2565}
\affiliation{%
  \institution{Booz Allen Hamilton}
  \city{McLean}
  \state{VA}
  \country{USA} 
}
\email{Zak_Richard@bah.com}

\author{Elliott Zaresky-Williams}
\orcid{0000-0003-4023-3880}
\affiliation{%
  \institution{Booz Allen Hamilton}
  \city{McLean}
  \state{VA}
  \country{USA} 
}
\email{Zaresky-Williams_Elliott@bah.com}

\author{Hyrum Anderson}
\orcid{0009-0009-4720-6907}
\affiliation{%
  \institution{Cisco Systems}
  \city{San Jose}
  \state{CA}
  \country{USA} 
}
\email{hyrum@cisco.com}

\author{Edward Raff}
\orcid{0000-0002-9900-1972}
\affiliation{%
  \institution{Booz Allen Hamilton}
  \city{McLean}
  \state{VA}
  \country{USA} 
}
\email{Raff\_Edward@bah.com}

\author{James Holt}
\orcid{0000-0002-6368-8696}
\affiliation{%
  \institution{Laboratory for Physical Sciences}
  \city{College Park}
  \state{MD}
  \country{USA} 
}
\email{holt@lps.umd.edu}

\renewcommand{\shortauthors}{Joyce et al.}

\begin{abstract}

A lack of accessible data has historically restricted malware analysis research, and practitioners have relied heavily on datasets provided by industry sources to advance. Existing public datasets are limited by narrow scope --- most include files targeting a single platform, have labels supporting just one type of malware classification task, and make no effort to capture the evasive files that make malware detection difficult in practice. We present EMBER2024, a new dataset that enables holistic evaluation of malware classifiers. Created in collaboration with the authors of EMBER2017 and EMBER2018, the EMBER2024 dataset includes hashes, metadata, feature vectors, and labels for more than 3.2 million files from six file formats. Our dataset supports the training and evaluation of machine learning models on seven malware classification tasks, including malware detection, malware family classification, and malware behavior identification. EMBER2024 is the first to include a collection of malicious files that initially went undetected by a set of antivirus products, creating a "challenge" set to assess classifier performance against evasive malware. This work also introduces EMBER feature version 3, with added support for several new feature types. We are releasing the EMBER2024 dataset to promote reproducibility and empower researchers in the pursuit of new malware research topics.

\end{abstract}


\maketitle

\section{Introduction}

Machine learning is increasingly being applied to several malware classification tasks \cite{raff2020survey}. Training and evaluating a malware classifier requires a large corpus of recently observed and well-labeled files, but sufficient data is not reasonably accessible to academics \cite{kantchelian}. Large security companies use deployed infrastructure and client telemetry to collect malware that is actively being distributed "in the wild" \cite{ugarte2019close}. Sharing agreements between such companies allow them to exchange files and threat intelligence. Commercial feeds of malware exist; however, subscribing to one may not be financially viable for an independent researcher \cite{virustotal, sorel}. This restricts access to newly emerging malware, especially if it is needed in large quantities. If a large dataset can be gathered, several challenges in publicly sharing both benign and malicious software often result in it being kept private \cite{saxe2015deep,mohaisen2015,huang2016mtnet}. These factors have made it difficult for researchers to compare the performance of their own malware classifiers against other work.

\begin{table}[!h]
\centering
\caption{Notable 1M+ file malware datasets. Our \textbf{EMBER2024} (E'24) is the only multi-platform dataset  to support malware detection, malware family classification, behavior identification. EMBER2024 is also the first to include a "challenge" set and code for replicating our dataset construction methodology.}
\label{tab:dataset_survey}
\begin{tabular}{@{}lccccc@{}}
\toprule
\multicolumn{1}{c}{} & SOREL      & MalDICT    & E'17     & E'18     & \textbf{E'24} \\ \midrule
Size                 & 20M        & 5.5M       & 1.0M       & 1.1M       & 3.2M              \\
Malw. Detection       & \Checkmark &            & \Checkmark & \Checkmark & \Checkmark        \\
Fam. Classification          &            &            &            & \Checkmark & \Checkmark        \\
Behav. Prediction        & \Checkmark & \Checkmark &            &            & \Checkmark        \\
Multi-Platform       &            &            &            &            & \Checkmark        \\
Challenge Set        &            &            &            &            & \Checkmark        \\
Infrastructure Code      &            &            &            &            & \Checkmark        \\ \bottomrule
\end{tabular}
\end{table}

To address this reproducibility issue, datasets for benchmarking malware classifiers have been released \cite{ember}. However, the most recent large (1M+ file) dataset with both benign and malicious software has files that were collected six years before the time of writing \cite{sorel}. The ecosystem of malware is constantly changing, and new varieties of malware, malicious techniques, and threat actors are not represented in prior datasets. This malware evolution results in concept drift, and performance degrades when a classifier attempts to detect malware that is newer than its training period \cite{jordaney}. Large datasets with recent malware exist, but they lack benign files, which are necessary for benchmarking malware detection \cite{maldict, virusshare}. Other benchmark datasets are limited to malware targeting a single platform and/or only have labels supporting a single classification task \cite{ember, ember2018, sorel}.

\subsection{The EMBER2024 Dataset}

With 3,238,315 files collected between September 2023 and December 2024, the EMBER2024 dataset provides researchers with a large and representative collection of recent malware. The files in EMBER2024 have \textbf{seven types of labels and tags} that support malware detection, family identification, and other relevant multi-label tasks. EMBER2024 is the first malware benchmark to include a \textbf{challenge set} of files that were not initially detected by any antivirus products. The dataset includes files in \textbf{six file formats} --- Win32, Win64, .NET, APK, ELF, and PDF --- and we introduce a feature format that allows unified representation and exploration across all formats. To our knowledge, we are also the first to provide code for replicating our dataset construction methodology. These properties aim to make EMBER2024 a holistic malware benchmark capable of assessing malware classifiers in a variety of ways.

The rest of this work is organized as follows. In Section \ref{sec:data-collection} we detail the design choices made in the creation of EMBER2024 and we discuss the process by which a set of antivirus-eluding malware was identified. Section \ref{dataset-contents} describes the contents of the EMBER2024 dataset and the updates made to the EMBER feature format, including partial support for non-PE files. Several experiments using standard benchmark models are performed in Section \ref{sec:benchmarks}, challenging conventional wisdom about classifier performance in the presence of evasive malware and concept drift. We present a retrospective written by the authors of EMBER2017 and EMBER2018 in Section \ref{sec:retrospective}, describing the impact of these datasets on the malware research community. Finally, we review related work in Section \ref{sec:related-work} and provide our conclusions in Section \ref{sec:conclusion}.

\section{Data Collection and Labeling}
\label{sec:data-collection}

In this section, we describe the procedure used to build the EMBER2024 dataset. On each day between September 24th, 2023 and December 14th, 2024, we identified a set of files that were first submitted to VirusTotal on that day. For each of those files, we retrieved VirusTotal analysis results (which we refer to as \textbf{VirusTotal reports}) within 24 hours of first submission. Then, we again queried each of those files 90 or more days after its first submission date. 

A VirusTotal report contains information about the queried file, such as its hashes, first submission date, last analysis date, antivirus (AV) detection results, and other metadata. A VirusTotal scan report for a fictitious file is shown in Figure \ref{fig:vtreport}. In the 90+ days between queries, files in VirusTotal may have been re-scanned, causing antivirus detections to update. We ensured that all files suspected to be benign received re-scans at least 30 days after first submission, re-scanning them ourselves if necessary. The most recent antivirus detections were then used to label and tag each file, described in more detail in Section \ref{sec:labeling-malware}. This methodology ensures that the malware in the EMBER2024 dataset is relevant and accurately labeled.

\lstset{
  backgroundcolor=\color{white},   
  basicstyle=\scriptsize\ttfamily, 
  breaklines=true,                 
  captionpos=b,                    
  frame=single,                    
  showstringspaces=false,          
  commentstyle=\color{gray},       
  keywordstyle=\color{blue},       
  stringstyle=\color{red},         
  numbers=none,                      
}
\begin{figure}[!t]

\begin{minted}[fontsize=\footnotesize,breaklines]{json}
{
  "sha256": "b7b78099082384d7da3594121d85dd7f4...",
  "first_submission_date": "2024-01-30T00:00:53",
  "last_analysis_date": "2024-04-06T11:52:27",
  "last_analysis_results": {
    "Microsoft": {
      "category": "malicious"
      "result": "TrojanDownloader:Win32/Nemucod!ml"
    },
    "MaxSecure": {
      "category": "malicious"
      "result": "Trojan.WIN32.cryxos.5913"
    },
    ...
  }
}
\end{minted}

\caption{Example VirusTotal report contents. VirusTotal reports include a file's hash, the date it was first submitted to VirusTotal, and the date the file was most recently analyzed. They also include scan results from $\approx$70 AV products.}
\label{fig:vtreport}
\end{figure}

\subsection{File Selection}

The file collection period for building the EMBER2024 dataset spans exactly 64 weeks. To encourage research into malware concept drift, EMBER2024 includes an equal number of files per week. Table \ref{tab:dataset_counts} lists how many files were included in the dataset from each week of data collection, per label and file format. The number of files per file type selected per week was determined by availability. For example, ELF files were the rarest of the six file types in our collection, and this is reflected in EMBER2024. 

\begin{table}[t!]
\centering
\vspace*{6pt}
\caption{EMBER2024 Weekly File Inclusion Thresholds}
\vspace*{-4pt}
\label{tab:dataset_counts}
\begin{tabular}{@{}lrrr@{}}
\toprule
File Type & Malicious Files & Benign Files 
 & Total \\
\midrule
Win32 & 15,000 & 15,000 & 30,000 \\
Win64 & 5,000 & 5,000 & 10,000 \\
.NET & 2,500 & 2,500 & 5,000 \\
APK & 2,000 & 2,000 & 4,000 \\
ELF & 250 & 250 & 500 \\
PDF & 500 & 500 & 1,000 \\
\bottomrule
\end{tabular}
\vspace*{-6pt}
\end{table}

For each of the 64 weeks of data collection, we gathered all files that we found to have a first VirusTotal submission date within that given week. Files from that week were then bucketed by file type and malicious-benign label. Any files that were not definitively identified as malicious or benign were discarded. Files were randomly drawn from each bucket (ignoring near-duplicates and files larger than 100MB in size) until the corresponding threshold in Table \ref{tab:dataset_counts} was reached. 

Near-duplicate files were identified using Trend Micro Locality Sensitive Hashing (TLSH) \cite{tlsh}. When considering whether to include a file in the dataset, we compared its TLSH digest against that of each other file already chosen from that week of data collection. If the current file had a TLSH distance of 30 or less to any previously selected file, the pair was considered to be near-duplicates, and the current file was discarded. This threshold was chosen based on the work of \citet{tlsh}, whose evaluation found that a TLSH distance of 30 has a false positive rate of 0.00181\% when identifying related files.

\subsection{Training and Test Sets}

EMBER2024 includes 50,500 files for each of the 64 weeks of data collection, divided into a training and test set. EMBER2024's training set is comprised of files from the first 52 weeks of data collection (Sep. 24, 2023 - Sep. 21, 2024, 2,626,000 files in total), while the final 12 weeks make up the test set (Sep. 22, 2024 - Dec. 14, 2024, 606,000 files in total).

\subsection{Challenge Set} The $\approx$70 AV products that VirusTotal uses for file scanning employ a variety of detection technologies, including file signatures, heuristics, and machine learning. File signatures are used to match known threats, while heuristics and machine learning attempt to detect emerging malware and malware that has changed enough to evade existing signatures \cite{botacin2020we}. In rare cases, malicious files go fully undetected on VirusTotal until AV products are updated with new signatures. Detecting evasive malware is a research priority, and EMBER2024 is the first dataset to provide a dedicated subset of files for evaluating classifiers on this task.

The 6,315 files in EMBER2024's "challenge" set were not initially detected by any AV products in VirusTotal. However, after being re-scanned at least 30 days later, they were detected by a sufficient number of AV products to receive a malicious label (described in Section \ref{sec:labeling-malware}). To maximize the size of the challenge set, files were selected from all 64 weeks of data collection (that is, overlapping the training and test set collection periods). Files in the challenge set do not appear in the EMBER2024 training or test sets. Furthermore, we ensured that for each file in the challenge set, no near-duplicates from the same week of data collection were included in the training or test sets. 

\begin{table}[b!]
\centering
\caption{EMBER2024 Dataset File Statistics}
\vspace*{-4pt}
\label{tab:dataset_splits}
\begin{tabular}{@{}lrrrr@{}}
\toprule
File Type & Train & Test & Challenge & Total \\
\midrule
Win32 & 1,560,000 & 360,000 & 3,225 & 1,923,225 \\
Win64 & 520,000 & 120,000 & 814 & 640,814 \\
.NET & 260,000 & 60,000 & 805 & 320,805 \\
APK & 208,000 & 48,000 & 256 & 256,256 \\
ELF & 26,000 & 6,000 & 386 & 32,386 \\
PDF & 52,000 & 12,000 &  805 & 12,805 \\
\bottomrule
\end{tabular}
\end{table}

\subsection{Labeling Methodology}
\label{sec:labeling-malware}
Like similar datasets, our labeling methodology is based on AV detection counts \cite{ember}. Files that were not detected by any AV products after being re-scanned at least 30 days after their first submission to VirusTotal are labeled as benign. Files detected as malware by five or more AV products without known relationships (e.g., engine sub-licensing, company acquisition, or public data sharing agreements) labeled as malicious \cite{maldict}.

The malware in EMBER2024 is labeled by family using ClarAVy \cite{claravy}. ClarAVy uses an intelligent Bayesian combination strategy to accurately predict malware families, and each family label has an associated confidence score. Every malicious file is also tagged according to its behaviors, file properties, exploits, packers, and threat group attribution using ClarAVy. Some files may not receive family labels or other tags due to insufficient information in their AV detections, or because no family or tag is applicable. Files may also receive multiple tags within the same category if applicable (for example, a file may have both the ``ransomware" and ``worm" behavioral tags). File property tags that indicate file format (e.g. ``win32", ``apk", ``pdf") were discarded due to redundancy.

\begin{figure}[b!]
    \centering
\vspace*{-6pt}
\includegraphics[width=0.95\columnwidth,keepaspectratio]{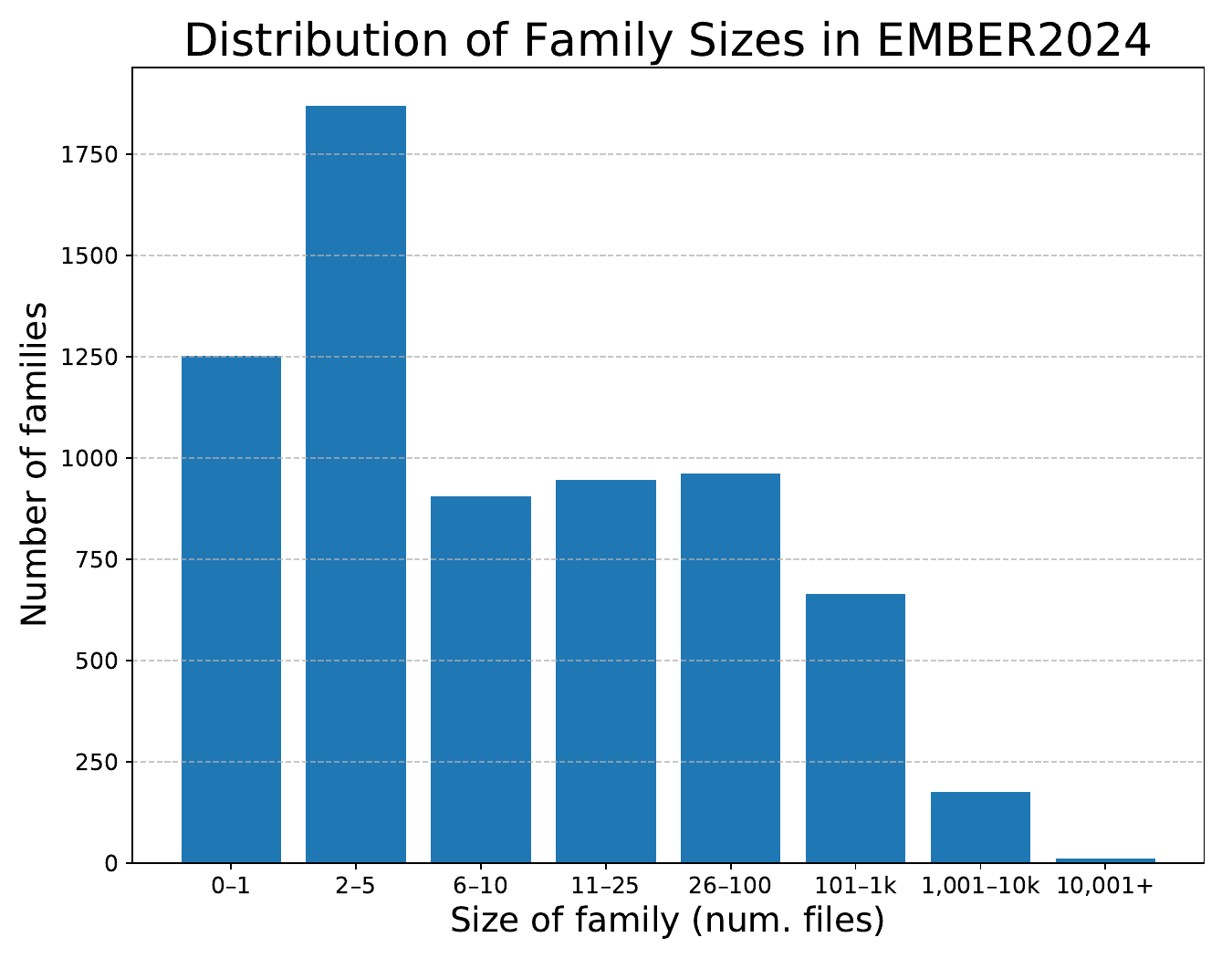}
    \vspace*{-8pt}
    \caption{Histogram showing the distribution of family sizes in the EMBER2024 training and test sets.}
    \label{fig:family-hist}

\end{figure}

\subsection{Family and Tag Demographics}
\label{sec:fam-demographics}

1,356,182 of the 1,616,000 malicious files in the EMBER2024 training and test sets have a malware family label. Consistent with prior work \cite{motif}, we observe that the malware family sizes in EMBER2024 approximately follow a zipfian distribution. A histogram of malware family sizes is shown in Figure \ref{fig:family-hist}. We identified 6,787 unique families among them, and 2,538 of those families have 10 or more instances in EMBER2024.  3,124 families have little representation in EMBER2024, with just five or fewer instances in the dataset. In contrast, there are 12 families that appear more than 10,000 times in EMBER2024. 34.04\% of the malicious files in the EMBER2024 training and test sets (550,087 files) belong to one of those 12 families. EMBER2024 has 2,709 files from 75 families that appear 10 or more times in the test set but are not in the training set. In section \ref{sec:eval-new-family} we use this to simulate the emergence of new families over time and evaluate the difficulty in detecting emerging families.

\begin{table}[t]
\centering
\vspace*{4pt}
\caption{EMBER2024 Tag Statistics}
\vspace*{-4pt}
\label{tab:tag-stats}
\begin{tabular}{@{}lrr@{}}
\toprule
Tag category & \# Tagged files & \# Total Tags \\
\midrule
Behavior & 733,142 & 118 \\
File property & 142,199 & 30 \\
Packer & 99,235 & 52 \\
Exploited vuln. & 2,991 & 293 \\
Threat group & 16,170 & 43\\
\bottomrule
\end{tabular}
\end{table}

Statistics about tags related to malware behaviors, file properties, packers, exploited vulnerabilities, and threat group attribution are listed in Table \ref{tab:tag-stats}. Tags that were applied to 10 or more files in EMBER2024 are used to train and evaluate benchmark classifiers in Section \ref{sec:multi-label}. Few files have tags related to vulnerability exploitation or known threat group.

\section{Dataset Contents}
\label{dataset-contents}

The contents of the EMBER2024 dataset include raw file metadata, feature vectors, labels and tags, and trained classifiers. Unfortunately, like EMBER2017 and EMBER2018, we cannot release the original files in EMBER2024. Instead, we are releasing code that allows users with a VirusTotal API key to download these files. We are also publishing other code related to dataset construction, feature extraction, and model training to aid researchers and ensure reproducibility.

\subsection{File Metadata and Feature Vectors}
\label{file-metadata}

The EMBER2024 dataset includes metadata in the form of JSON objects for each file. Files are uniquely identified by MD5, SHA-1, and SHA-256 digests, and TLSH digests enable approximate file comparison. Each JSON object includes UNIX timestamps indicating when a file was first uploaded to VirusTotal and when the file was most recently scanned in VirusTotal (at the time of dataset construction). Each object also lists the ratio of malicious detections to total antivirus scans, the file type, family label, and other tags. Finally, each JSON object includes EMBER feature version 3 raw features for the corresponding file. An JSON object for a fictitious file is shown in Figure \ref{fig:report_json}.

\subsection{EMBER Feature Version 3}
\label{sec:emberv3}

The EMBER2017 and EMBER2018 datasets established version 1 and version 2 of the EMBER feature vector format, respectively \cite{ember, ember2018}. The EMBER feature format enables researchers to easily obtain raw features and/or feature vectors from Windows Portable Executable (PE) files, and it has been broadly adopted by the malware analysis research community for training classifiers to perform static detection of Windows malware 
\cite{galen2020evaluating,
loi2021towards, 8858297, song2021mabmalwarereinforcementlearningframework}. We introduce EMBER feature version 3, which updates existing feature categories from feature versions 1 and 2 and adds several new feature categories. Furthermore, feature version 3 supports limited feature extraction from non-PE files and PE files that cannot be properly parsed.  

EMBER feature version 3 includes features extracted from the PE COFF, Optional, and Section headers that were not included in feature versions 1 or 2. Modifications were made to the set of general file features and more features related to string appearances and statistics have been incorporated. Several new feature categories were added to EMBER feature version 3: DOS header features, data directory features, Rich header features, Authenticode signature features, and PE file parse warning features. Appendix \ref{sec:appendixA} displays EMBER version 3 raw features extracted from a fictitious file. We discuss each of the feature version 3 changes below.

\textbf{Additonal PE COFF Header Features.} Four features from the PE COFF header were added: The number of PE sections, the size of the PE optional header, the number of symbols, and the pointer to the symbol table.

\textbf{Additional PE Optional Header Features.} 13 features from the PE Optional header were added to EMBER feature version 3. These features are primarily the sizes of, or pointers to, various regions of memory required by the Windows loader when a PE file is executed and becomes a process.

\textbf{Additional PE Section header features.} Feature version 3 reduces the number of features allocated to PE section names. Features such as the ratio of physical section size to total file size and the ratio of physical section size to virtual section size, have been added.

\textbf{General File Features.} The general file feature category has been repurposed, and now includes file size, file entropy, and the first four bytes in the file (for inferring file type). Features specific to Windows PE files have been moved to other feature categories.

\textbf{String Features.} Prior EMBER feature versions included regular expressions that search for strings indicative of file paths, URLs, registry keys, and embedded PE executables. Feature version 3 now searches for 76 strings or string patterns that may be useful for determining a file's behaviors and whether it is malicious or benign.

\textbf{DOS Header Features.} The DOS header remains at the beginning of PE files for legacy purposes. EMBER feature version 3 includes each entry in the DOS header as a feature.

\lstset{
  backgroundcolor=\color{white},   
  basicstyle=\scriptsize\ttfamily, 
  breaklines=true,                 
  captionpos=b,                    
  frame=single,                    
  showstringspaces=false,          
  commentstyle=\color{gray},       
  keywordstyle=\color{blue},       
  stringstyle=\color{red},         
  numbers=none,                      
}
\begin{figure}[b]
\vspace*{6pt}
\begin{lstlisting}
{
  "md5": "93080b69b30c4658ecaf4104f8bf62d5"
  "sha1": "14bb95b5220acb12c328922567cc899330e...",
  "sha256": "b7b78099082384d7da3594121d85dd7f4...",
  "tlsh": "T1002354D8E1FEDE31036602DDB3E9AB5B7...",
  "first_submission_date": 1704706843,
  "last_analysis_date": 1707870639,
  "detection_ratio": "64/75"
  "label": 1,
  "file_type": "Win32",
  "family": "wannacry",
  "family_confidence": 0.961,
  "behavior": ["ransomware", "worm"],
  "file_property": ["packed"],
  "packer": ["upx"],
  "exploit": ["cve-2017-0144],
  "group": [],
  "histogram": [...],
  "byteentropy": [...],
  "strings: {...},
  "general": {...},
  "header": {...},
  "section": {...},
  "imports": {...},
  "exports": [...],
  "datadirectories": {},
  "richheader": [...],
  "authenticode": {...},
  "pefilewarnings": [...],
}
\end{lstlisting}
\caption{Example JSON object displaying a file's hashes, labels, EMBER feature version 3 raw features, and other metadata.}
\vspace*{-4pt}
\label{fig:report_json}
\end{figure}

\textbf{PE Data Directory Features.} A PE file may contain several data directories with specialized information, such as debug data, relocation data, and resource data. The names, sizes, and virtual addresses of each data directory are used as features.

\textbf{Rich Header Features.} The Rich header is an undocumented header included in PE files linked using the Windows loader. It includes metadata about artifacts generated during the compilation and linking process. EMBER feature version 3 uses the hashing trick to record each entry in the Rich header.

\textbf{Authenticode Signature Features.} Authenticode is used for Windows PE file code signing. EMBER feature version 3 includes features about Authenticode signatures such as the number of certificates, whether the certificate is self-signed, whether any certificates have empty name values, the date of the most recent certificate, and the difference between this date and the file's compilation timestamp.

\textbf{PE Parse Warning Features.} The \texttt{pefile} library (now used for extracting many EMBER raw features) may throw several errors and/or warnings when parsing a file. This is relevant since many malicious PE files --- especially those that have been packed or modified after compilation --- may not parse correctly. EMBER feature version 3 includes 88 features for tracking various errors and warnings during file parsing.

EMBER feature version 3 is able to partially handle non-PE files and PE files that cannot be parsed. In these instances, the following feature categories can still be extracted: general file features, string patterns and statistics, byte histogram features, and byte entropy histogram features. In Section \ref{sec:benchmarks} we show that this limited feature set remains effective for classifying Linux, Android, and PDF malware.

The EMBER feature version 3 raw features vectorize into a feature vector of dimension 2,568 (previously 2,381 under feature version 2). EMBER vectors for non-PE files can be safely truncated to dimension 696, since all further entries in those vectors are zero.

\subsection{Source Code}
\label{sec:source-code}

We recognize that, like prior work, the EMBER2024 dataset will become outdated over time. We are publishing code for the following:

\begin{itemize}
    \item Retrieving VirusTotal reports for a collection of files.
    \item Computing TLSH digests for a collection of files.
    \item Labeling a collection of files as malicious or benign using the antivirus results in VirusTotal reports.
    \item Selecting a preset number of files from a collection, with near-duplicates excluded.
    \item Downloading selected files from VirusTotal.
\end{itemize}

This code will allow researchers to replicate the methodology used to compile the EMBER2024 dataset. Note that this requires the use of VirusTotal's API. We are also releasing a Python code implementing the following:

\begin{itemize}
    \item Extracting EMBER feature version 3 raw metadata.
    \item Vectorizing EMBER raw features, and writing feature vectors and labels/tags to disk.
    \item Reading EMBER feature vectors and labels/tags from disk.
    \item Training \texttt{LightGBM} classifiers on EMBER feature vectors.
\end{itemize}

This released code is an update to the original EMBER Python package. The EMBER feature version 1 and 2 implementations in the original package use the \texttt{LIEF} library to extract PE metadata features \cite{LIEF}. However, these implementations are pinned to older versions of \texttt{LIEF} and have outdated dependencies. Over time, installing these versions of \texttt{LIEF} to compute EMBER raw features has become more burdensome. To rectify this, the implementation of the EMBER feature version 3 implementation has switched from \texttt{LEIF} to \texttt{pefile}, a well-supported and robust library that has no other dependencies \cite{pefile}. 

We have also added code functionality that enables users to load a subset of the EMBER2024 feature vectors for their required classifier training and/or evaluation task(s). Users can easily create custom dataset splits from any combination of the following:

\begin{itemize}
    \item The training set, test set, or challenge set.
    \item PE files, Win32 files, Win64 files, .NET files, APK files, ELF files, or PDF files.
    \item Files with malicious-benign labels (i.e. all files), files with family labels, or files with a certain type of tag.
\end{itemize}

\section{Benchmark Model Results}
\label{sec:benchmarks}

We are releasing 14 \texttt{LightGBM} classifiers trained to perform various malware analysis classification tasks including malware detection, malware family identification, and malware attribute prediction. This section includes training details, benchmark results, and discussion of our findings.

\begin{figure}[H]
    \centering

\includegraphics[width=0.93\columnwidth,keepaspectratio]{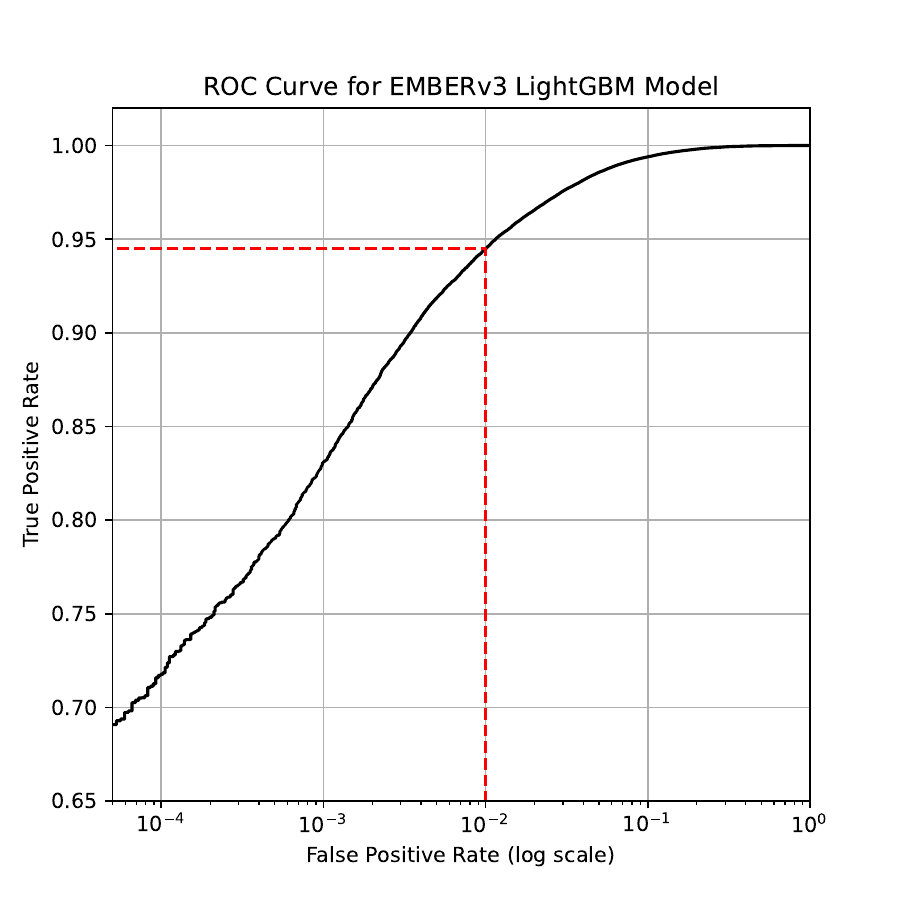}
    \caption{ROC curve (log scale) of a \texttt{LightGBM} classifier trained on the EMBER2024 training set and evaluated using the EMBER2024 test set. The model has a true positive rate of 94.48\% when permitting a 1\% false positive rate.}
    \label{fig:classifier-results}
\end{figure}

\subsection{Evaluating a Malware Detection Classifier}
\label{sec:all-files-classifier}

We trained a \texttt{LightGBM} classifier on the EMBER2024 training set for 500 boosting rounds, with 64 leaves and 100 minimum data points required per leaf. These hyper-parameters are identical to the SOREL dataset's benchmark \texttt{LightGBM} classifier \cite{sorel}. Figure \ref{fig:classifier-results} shows the classifier's ROC curve for the EMBER2024 test set. The \texttt{LightGBM} model has a ROC AUC score of 0.9949 and a true positive rate of 94.48\% at 1\% false positive rate.

\begin{table}[b!]
\centering
\caption{ROC AUC and PR AUC scores for \texttt{LightGBM} classifiers trained on various partitions of the EMBER2024 training set and evaluated using different parts of the EMBER2024 test set.}
\vspace*{-4pt}
\label{tab:filetype-eval-test}
\begin{tabular}{@{}llrr@{}}
\toprule
Training partition & Evaluation partition & ROC AUC & PR AUC \\ 
\midrule
All files & All files & 0.9969 & 0.9971 \\
          & All PE files & 0.9982 & 0.9983 \\
          & Win32 files & 0.9981 & 0.9983 \\
          & Win64 files & 0.9983 & 0.9985 \\
          & .NET files & 0.9968 & 0.9968 \\
          & APK Files & 0.9726 & 0.9737 \\
          & ELF files & 0.9887 & 0.9902 \\
          & PDF files & 0.9878 & 0.9901 \\
\midrule
All PE files & All PE files & 0.9982 & 0.9983 \\
             & Win32 files & 0.9982 & 0.9984 \\
             & Win64 files & 0.9986 & 0.9987 \\
             & .NET files & 0.9971 & 0.9971 \\
\midrule
Win32 files & Win32 files & 0.9984 & 0.9986  \\
Win64 files & Win64 files & 0.9989 & 0.9990 \\
.NET files & .NET files & 0.9980 & 0.9981 \\
APK files & APK files & 0.9868 & 0.9877 \\
ELF files & ELF files & 0.9933 & 0.9933 \\
PDF files & PDF files & 0.9912 & 0.9933 \\

\bottomrule
\end{tabular}
\end{table}

\subsection{Evaluating EMBER2024 Subset Classifiers}
\label{sec:file-classifiers}

Eight \texttt{LightGBM} classifiers listed in Table \ref{tab:filetype-eval-test} were trained on the following partitions of the EMBER2024 training set. The "All files" classifier was trained on the entire EMBER2024 training set, and the "All PE files" classifier was trained using the Win32, Win64, and .NET files in the training set. The "All files" and "All PE files" classifiers were then evaluated several times, using different subsets of the EMBER2024 test set. Classifiers for Win32 files, Win64 files, .NET files, APK files, ELF files, and PDF files were also trained and evaluated using appropriate partitions of EMBER2024. All classifiers use the same hyperparameters described in Section \ref{sec:all-files-classifier}.

The ROC AUC and Precision-Recall (PR) AUC scores in Table \ref{tab:filetype-eval-test} indicate that the trained \texttt{LightGBM} classifiers are able to accurately detect malicious files in the EMBER2024 test set, which consists of files that appeared in VirusTotal for the first time 1-12 weeks after the most recent file in the training set. This suggests that the EMBER feature version 3 features are resistant to concept drift due to small temporal changes.

The ROC AUC and PR AUC scores of the ELF and PDF classifiers in Table \ref{tab:filetype-eval-test} exceed 0.99, despite being trained on a reduced set of features. The APK classifier was also trained with this limited feature set, and although it was outperformed by the ELF and PDF classifiers, its ROC AUC and PR AUC scores of 0.9868 and 0.9877 demonstrate that it can effectively detect APK malware.

\begin{table}[b!]
\centering
\caption{ROC AUC and PR AUC scores for \texttt{LightGBM} classifiers, using malicious files in the EMBER2024 challenge set plus benign files in the EMBER2024 test set.}
\vspace*{-4pt}
\label{tab:filetype-eval-challenge}
\begin{tabular}{@{}llrr@{}}
\toprule
Training partition & Evaluation partition & ROC AUC & PR AUC \\ 
\midrule
All files & All files & 0.9542 & 0.5722 \\
          & All PE files & 0.9643 & 0.6250 \\
          & Win32 files & 0.9662 & 0.6526 \\
          & Win64 files & 0.9424 & 0.3214 \\
          & .NET files & 0.9773 & 0.7940 \\
          & APK files & 0.8700 & 0.7644 \\
          & ELF files & 0.8975 & 0.5200 \\
          & PDF files & 0.7804 & 0.6145 \\
\midrule
All PE files & All PE files & 0.9661 & 0.6354 \\
             & Win32 files & 0.9675 & 0.6540 \\
             & Win64 files & 0.9464 & 0.3509 \\
             & .NET files & 0.9790 & 0.8066  \\
\midrule
Win32 files & Win32 files & 0.9689 & 0.6646 \\
Win64 files & Win64 files & 0.9503 & 0.4651 \\
.NET files & .NET files & 0.9865 & 0.8539 \\
APK files & APK files & 0.9101 & 0.2374 \\
ELF files & ELF files & 0.9311 & 0.6008 \\
PDF files & PDF files & 0.9066 & 0.7841 \\

\bottomrule
\end{tabular}
\end{table}

\subsection{Evaluations Using the Challenge Set}
\label{sec:challenge-classifiers}

Next, we repeated the experiments in Section \ref{sec:file-classifiers} using the EMBER2024 challenge set, with results reported in Table \ref{tab:filetype-eval-challenge}. Because the challenge set contains only malware, each classifier was evaluated by joining a partition of the challenge set with the benign files in its corresponding partition of the test set. For example, the "PDF files" entry in the "Evaluation partition" column of Table \ref{tab:filetype-eval-challenge} refers to the 805 malicious PDF files in the challenge set plus the 6,000 benign PDF files in the test set. 

The ROC AUC scores in Table \ref{tab:filetype-eval-challenge} do not accurately reflect model performance due to class imbalance. PR AUC scores demonstrate that \texttt{LightGBM} classifiers struggle to detect the malware in the challenge set, just as AV products initially did. The .NET classifier was able to most accurately identify challenge files, with a PR AUC score of 0.8539. APK files and Win64 files in the challenge set were particularly difficult to detect, with their classifiers having PR AUC scores of 0.2374 and 0.4651 respectively.

Many malware detection tasks require extremely low false positive rates (e.g. below 0.1\%), necessitating large evaluation sets to differentiate model performance \cite{sorel, patel2023small}. The difficulty of the challenge set allows researchers to quickly estimate the relative performance of malware classifiers using a small-scale dataset. Our experiments using the challenge set establish a benchmark for detecting evasive malware and demonstrate that there is potential for significant improvement in this research area.

\subsection{Detecting Newly Emerging Families}
\label{sec:eval-new-family}

Recall that the EMBER2024 test set includes 2,709 files from 75 families of size 10 or greater that do not appear in the training set. These files simulate "newly emerging" families for which signatures may not yet exist, and prior work indicates that these files may be more difficult to detect \cite{patel2023small}. We combined these 2,709 malicious files with the 303,000 benign files in the EMBER2024 test set and evaluated the \texttt{LightGBM} malware detection classifier from Section \ref{sec:all-files-classifier} using them. The classifier's resulting PR AUC score was 0.8992.

Our findings in Section \ref{sec:file-classifiers} imply that our baseline \texttt{LightGBM} classifiers are resistant to multiple weeks of concept drift. During such a period, new versions of existing families are introduced and entirely novel families emerge. It seems that our benchmark \texttt{LightGBM} classifiers are well-equipped to identify derivative versions of existing families. However, performance clearly degrades for detection of novel families, and we encourage more study on this topic.

\subsection{Evaluating Malware Family Classification}
\label{sec:family-classifier}

Next, we trained a \texttt{LightGBM} classifier to perform malware family identification. We identified 2,358 families that appear 10 or more times in the EMBER2024 training set. Files not in one of these 2,358 families were not used for model training. The remaining files were divided using a stratified split, with 90\% of each family used for training and 10\% for validation during training. The \texttt{LightGBM} classifier was trained for 100 boosting rounds, with 64 leaves and 10 minimum data points per leaf. Early stopping was permitted after 10 boosting rounds. Evaluation was performed using all files with family labels in the EMBER2024 test set, and the classifier achieved an accuracy of 67.97\%. We also computed the precision, recall, and F1 score of the classifier using both macro averaging and weighed averaging, and these results are shown in Table \ref{tab:family-eval}.

The model's performance metrics are markedly lower when using macro averaging. This suggests that the model performs well in detecting common families, but has more difficulty classifying the (many) smaller families in the dataset. We believe that this is primarily due to lack of sufficient training data for these smaller families.

\subsection{Multi-Label Malware Classification Tasks}
\label{sec:multi-label}

\texttt{LightGBM} classifiers were trained to perform the following multi-label tasks: behavior prediction, file property prediction, packer identification, exploited vulnerability identification, and threat group identification. Classifiers were trained on individual tags using a One-Vs-Rest (OvR) approach, using the same hyper-parameters as the \texttt{LightGBM} model in Section \ref{sec:family-classifier}. Tags that occurred fewer than 10 times in the EMBER2024 training set were discarded from the training and test sets.

The results in Table \ref{tab:task-evaluation-test} show that the classifiers clearly struggled to generalize in all five of these tasks, with low results across all metrics. The precision of models with a large number of tags was especially low. Like family classification, the poor macro-averaging results are likely due to limited training data for many tags, in addition to the difficulty of each task. Prior work also points to a temporal train/test split contributing to lowered performance in multi-label malware classification \cite{maldict}.

\subsection{Discussion of Benchmark Results}

The \texttt{LightGBM} classifiers used in our experiments have not been tuned and are meant to leave room for optimization. Improvements in hyper-parameter selection, model choice, and training strategy will likely yield better performance. Rather, these models are meant to serve as benchmarks that demonstrate the results that can be expected from a basic classifier trained on these tasks. EMBER2024 enables researchers to publish reproducible results by evaluating their classifiers against our benchmark models and models trained by others.

The \texttt{LightGBM} benchmark classifiers of EMBER2018 and SOREL-20M have ROC AUC scores of 0.996 and 0.998, respectively, despite attempts to include more "difficult" malware than EMBER2017 \cite{ember2018, sorel, ember}. However, our studies on evasive malware and malware from novel families show that malware detection is far from a solved problem. Despite our own \texttt{LightGBM} classifier having a ROC AUC score (0.9969) similar to past benchmarks, we identified populations of malicious files that can reliably bypass detection. We believe that further studies in this area are warranted, and the EMBER2024 test and challenge sets will support this research.

\begin{table}[t]
\centering
\caption{Family classification results for a \texttt{LightGBM} model trained to identify 2,358 families.}
\vspace*{-4pt}
\label{tab:family-eval}
\begin{tabular}{@{}lrr@{}}
\toprule
Metric & Score (macro avg.) & Score (weighted avg.)\\ 
\midrule
Precision & 0.5670 & 0.7360 \\
Recall & 0.3980 & 0.6797\\
F1 score & 0.4371 & 0.6664\\
\bottomrule
\end{tabular}
\end{table}

\begin{table}[t]
\centering
\caption{Precision, Recall, F1 Measure, and average AUC (using macro averaging) of One-vs-Rest (OvR) \texttt{LightGBM} classifiers trained to predict tags in EMBER2024.}
\vspace*{-4pt}
\label{tab:task-evaluation-test}
\begin{tabular}{@{}lrrrrr@{}}
\toprule
Pred. Task & \# Tags & Precision & Recall & F1 & AUC \\ 
\midrule
Behavior & 92 & 0.0981 & 0.5254 & 0.1345 & 0.7558\\
File property & 20 & 0.3037 & 0.5328 & 0.3451 & 0.7462 \\
Packer  & 32 & 0.2066 & 0.6722 & 0.2525 & 0.8310 \\
Exploited vuln.  & 46 & 0.5038 & 0.6570 & 0.5102 & 0.8192\\
Threat group & 6 & 0.7588 & 0.5488 & 0.5823 & 0.7737\\
\bottomrule
\end{tabular}
\end{table}

\section{EMBER Dataset Retrospective}
\label{sec:retrospective}
Previous to the release of the first EMBER dataset and accompanying \texttt{LightGBM} model, several pioneering works applied machine learning to train malware classifiers \cite{cohen1995fast,schultz2001data,kolter2004learning,saxe2015deep}; however, datasets were proprietary and/or very small (a few thousand samples).  Initially released under the generous support of Endgame (now part of Elastic) in 2018, EMBER was created with a straightforward goal: to provide a standardized benchmark dataset to ``invigorate machine learning research for malware detection'' in much the same way that benchmark datasets had done for computer vision \cite{ember}. We considered a number of research use-cases that included baselining malware classification performance with the co-released \texttt{LightGBM} model,
adversarial machine learning offense and defense, semi-supervised learning for malware detection, among others.

Since its release, the original EMBER dataset has been cited over 600 times from more than 350 unique citing institutions across 6 continents, in what we considered to be a relatively niche research field at the time.  A sampling of papers shows an 82\% / 18\% split between academia / industry affiliations.  A brief survey of citing publications indicate that the EMBER publication has also spurred the release of other malware datasets, in addition to innovations in defensive ML security (e.g., malware classification), offensive ML security (e.g., malware evasion), and advancements in ML architectures or algorithms.

\begin{table}[H]
\centering
\caption{Topics of papers that cited the first EMBER dataset, as adjudicated with the assistance of GPT-4o.}

\label{tab:topics}
\begin{tabular}{@{}lr@{}}
\toprule
Category & Percent \\
\midrule
Defensive ML Security  & 36.2\%\\
Survey Papers & 19.8\% \\
Other Benchmark Datasets & 19.0\% \\
Offensive ML Security & 18.1\%\\
ML Architecture or Algorithms & 6.9\% \\
\bottomrule
\end{tabular}
\end{table}

Besides academic publications, a host of unpublished work from malware offensive and defensive competitions \cite{Apruzzese2023,MicrosoftSecurityTeam2021} has engaged security practitioners. Email interactions with educators indicate that the dataset and model are being used at institutions that range from high school to graduate school (and no, we are still legally unable to provide benign files).

In summary, we have been overwhelmed by the response to the EMBER dataset. This overdue update will make the EMBER features easier to calculate, includes more capable features for Windows PE files, expands support beyond Windows PE files, and will enable yet another generation of researchers to advance the state of the art for applying machine learning to malware detection and related challenges.

\section{Related Work}
\label{sec:related-work}

Following EMBER2017 and EMBER2018, other 1M+ file malware datasets have contributed to the malware research domain. The SOREL-20M dataset was the first large, labeled dataset to release disarmed malicious executables \cite{sorel}, and it is currently the largest dataset with labeled malicious and benign files at the time of writing. Furthermore, the malware in SOREL-20M is tagged according to 11 behavioral properties. MalDICT is a malware-only dataset tagged according to malicious behaviors, file properties, exploited vulnerabilities, and file packers \cite{maldict}. It made benchmarking less common malware classification tasks possible for the first time. The VirusShare collection is the largest public malware corpus to our knowledge, with 41,680,896 files available for download at the time of writing. AVClass labels for $\approx$79.5\% of these files (from April 2019 and earlier) are available online \cite{seymour, avclass}. VirusShare is regularly updated with new malware, but does not include benign files.

\section{Conclusion}\label{sec:conclusion}

The malware research community is long overdue for another large malware benchmark dataset. EMBER2024 gives researchers access to metadata, feature vectors, and labels for more than 3.2 million malicious and benign files. Including six file formats and seven types of labels and tags, our dataset makes holistic evaluation of malware classifiers attainable. To our knowledge, the EMBER2024 challenge set is the first of its kind, enabling new studies on evasive malware. As a result of our inquiries into evasive malware and newly emerging families, we advocate for further study on developing robust malware classifiers. We also made several other contributions in this work, such as code for replicating our dataset building methodology, an updated EMBER version 3 feature vector format, and 14 trained benchmark classifiers. The main EMBER2024 GitHub repository can be found at \href{https://github.com/FutureComputing4AI/EMBER2024}{https://github.com/FutureComputing4AI/EMBER2024}. Code for replicating our dataset building methodology is located at \href{https://github.com/FutureComputing4AI/vtpipeline-rs}{https://github.com/FutureComputing4AI/vtpipeline-rs}. It is our hope that EMBER2024 will become a valuable resource for researchers and a catalyst for investigating critical malware analysis topics.

\bibliographystyle{ACM-Reference-Format}

\clearpage
\appendix

\section{EMBER Feature Version 3 Raw Features}
\label{sec:appendixA}

\lstset{
  backgroundcolor=\color{white},   
  basicstyle=\scriptsize\ttfamily, 
  breaklines=true,                 
  captionpos=b,                    
  frame=single,                    
  showstringspaces=false,          
  commentstyle=\color{gray},       
  keywordstyle=\color{blue},       
  stringstyle=\color{red},         
  numbers=none,                      
}
\begin{figure}[!htb]

\begin{minted}[fontsize=\footnotesize,breaklines]{json}
{
  "histogram": [67647, 42400, 37862, ..., 32387, 33015, 37394],
  "byteentropy": [0, 0, 0, 0, ..., 1058323, 1063221, 1051062],
  "strings": {
    "numstrings": 43473,
    "avlength": 6.213189795965312,
    "printabledist": [3321, 2643, 2944, 2692, 3016, ...],
    "printables": 270106,
    "entropy": 6.582472077598922,
    "string_counts": {
      "btc_wallet": 1,
      "certificate": 8, 
      "connect": 11, 
      "crypt": 31
      ...
    }
  },
  "general": {
    "size": 8782336,
    "vsize": 8880128,
    "has_relocs": 1,
    "has_dynamic_relocs": 0,
    "symbols": 0
  },
  "header": {
    "coff": {
      "timestamp": 1695592800,
      "machine": "IMAGE_FILE_MACHINE_AMD64",
      "number_of_sections": 12,
      "number_of_symbols": 0,
      "sizeof_optional_header": 240,
      "pointer_to_symbol_table": 0,
      "characteristics": ["EXECUTABLE_IMAGE", ...],
    },
    "optional": {
      "magic": 523,
      "subsystem": "IMAGE_SUBSYSTEM_WINDOWS_CUI",
      "major_image_version": 0,
      "minor_image_version": 0,
      "major_linker_version": 2,
      "minor_linker_version": 38,
      "major_operating_system_version": 4,
      "minor_operating_system_version": 0,
      "major_subsystem_version": 5,
      "minor_subsystem_version": 2,
      "sizeof_code": 115200,
      "sizeof_headers": 1024,
      "sizeof_image": 8880128,
      "sizeof_initialized_data": 8781312,
      "sizeof_uninitialized_data": 65024,
      "sizeof_stack_reserve": 2097152,
      "sizeof_stack_commit": 4096,
      "sizeof_heap_reserve": 1048576,
      "sizeof_heap_commit": 4096,
      "address_of_entrypoint": 4389,
      "base_of_code": 4096,
      "base_of_data": 0,
      "image_base": 5368709120,
      "section_alignment": 4096,
      "checksum": 184607,
      "number_of_rvas_and_sizes": 16,
      "dll_characteristics": ["HIGH_ENTROPY_VA", ...],
    },
    "dos": {
      "e_magic": 23117,
      "e_cblp": 144,
      "e_cp": 3,
      "e_crlc": 0,
      "e_cparhdr": 4,
      ...
\end{minted}
\end{figure}

\newpage

\lstset{
  backgroundcolor=\color{white},   
  basicstyle=\scriptsize\ttfamily, 
  breaklines=true,                 
  captionpos=b,                    
  frame=single,                    
  showstringspaces=false,          
  commentstyle=\color{gray},       
  keywordstyle=\color{blue},       
  stringstyle=\color{red},         
  numbers=none,                      
}


\begin{figure}[!htb]
\vspace*{23.5pt}
\begin{minted}[fontsize=\footnotesize,breaklines]{json}
      "e_minalloc": 0,
      "e_maxalloc": 65535,
      "e_ss": 0,
      "e_sp": 184,
      "e_csum": 0,
      "e_ip": 0,
      "e_cs": 0,
      "e_lfarlc": 64,
      "e_ovno": 0,
      "e_oemid": 0,
      "e_oeminfo": 0,
      "e_lfanew": 128
    }
  },
  "section": {
    "entry": ".text",
    "sections": [
      {
        "name": ".text",
        "size": 115200,
        "entropy": 6.2928493046865155,
        "vsize": 115080,
        "size_ratio": 0.013117238966944557,
        "vsize_ratio": 1.0010427528675705,
        "props": ["CNT_CODE", ...],
      },
      ...
    ],
    "overlay": {
      "size": 6935696,
      "size_ratio": 0.9556944024939537,
      "entropy": 7.997956389085634
    }
  },
  "imports": {
    "KERNEL32.dll": ["CloseHandle", "CopyFileW", ...],
    "SHELL32.dll": ["SHFileOperationW", ...],
    ...
  },
  "exports": [],
  "datadirectories": [
    {
      "name": "RESOURCE",
      "size": 8643008,
      "virtual_address": 229376
    },
    {
      "name": "IAT",
      "size": 768,
      "virtual_address": 217936
    },
    ...
  ],
  "richheader": [1704619, 7, 17135691, 191, 170705, ...],
  "authenticode": {
    "num_certs": 2,
    "self_signed": 1,
    "empty_program_name": 0,
    "no_countersigner": 0,
    "parse_error": 0,
    "chain_max_depth": 7,
    "latest_signing_time": 1643104921,
    "signing_time_diff": 19080976
  },
  "pefilewarnings": [
    "RVA AddressOfFunctions in the export directory...",
    "Invalid IMAGE_DYNAMIC_RELOCATION_TABLE...",
    ...
  ]
}
\end{minted}
\vspace*{-4pt}
\caption{Example of EMBER feature version 3 raw features.}
\label{fig:emberv3_features}
\end{figure}

\newpage
\onecolumn

\section{Top Families and Tags in EMBER2024}

\begin{figure}[h!]
    \centering
\vspace*{-6pt}
\includegraphics[width=0.75\columnwidth,keepaspectratio]{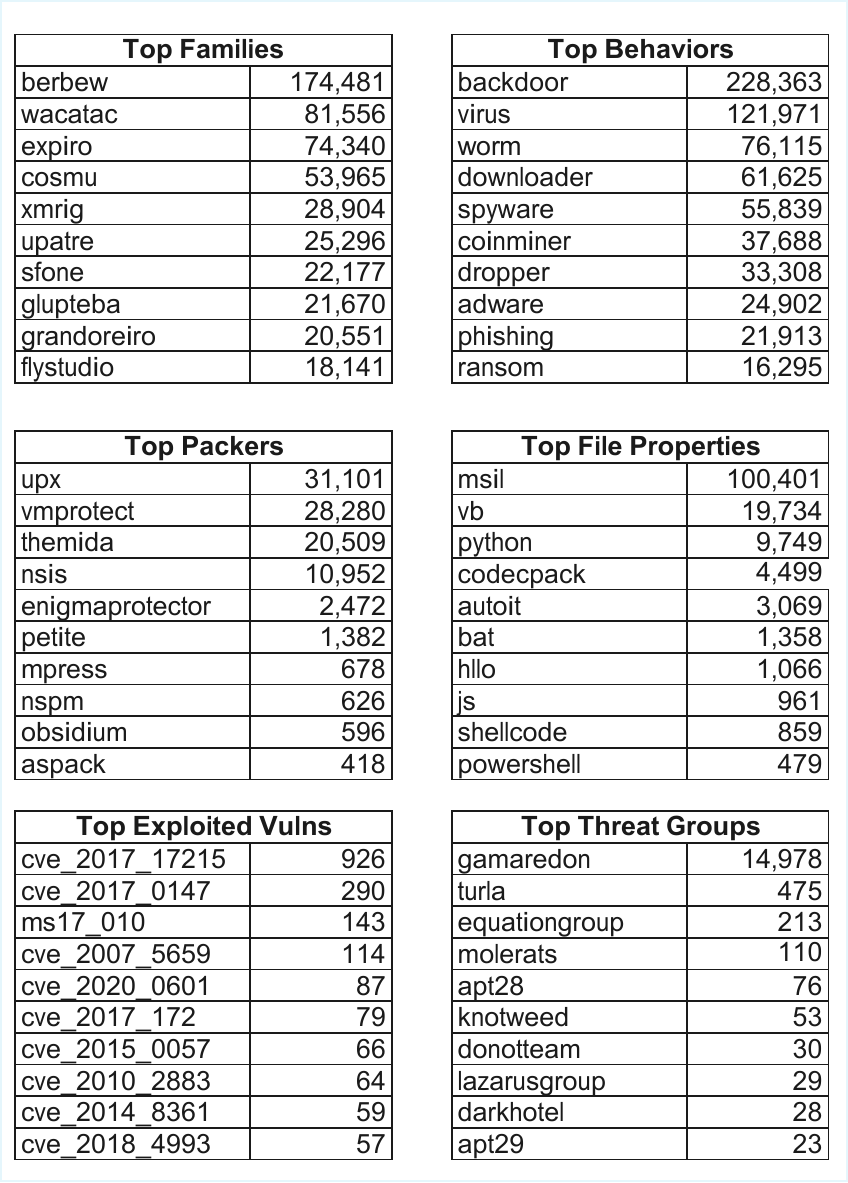}
    \vspace*{-8pt}
    \caption{Most common families and tags in the EMBER2024 dataset.}
    \label{fig:top-tags}

\end{figure}

\end{document}